\documentclass{elsart}

\usepackage{graphicx}

\usepackage{amssymb}

\begin{document}

\begin{frontmatter}



\title{Evolutionary Construction of Geographical Networks with 
Nearly Optimal Robustness and Efficient Routing Properties} 


\author[JAIST]{Yukio Hayashi}

\address[JAIST]{Japan Advanced Institute of Science and Technology, 
Ishikawa 923-1292, Japan}

\begin{abstract}
Robust and efficient design of networks
on a realistic geographical space
is one of the important issues
for the realization of dependable communication systems.
In this paper,
based on a percolation theory and a geometric graph property,
we investigate such a design from the following viewpoints:
1) network evolution according to a spatially heterogeneous
 population,
2) trimodal low degrees for the tolerant connectivity
against both failures and attacks,
and
3) decentralized routing within short paths.
Furthermore, we point out
the weakened tolerance by geographical constraints on local cycles,
and propose a practical strategy by adding 
a small fraction of shortcut links between randomly chosen nodes
in order to improve the robustness to a similar level to that of 
the optimal bimodal networks with a larger degree $O(\sqrt{N})$ 
for the network size $N$. 
These properties will be useful for
constructing future ad-hoc networks in wide-area communications.
\end{abstract}

\end{frontmatter}

\section{Introduction} \label{sec1}
Real complex networks,
such as a power grid, an airline network, and the Internet,
are embedded in a metric space,
and long-range links are restricted \cite{Yook02,Gastner06}
for economical reasons. 
Moreover, 
there exists a common topological characteristic called 
{\em scale-free}(SF) that follows a power-law
degree distribution $P(k) \sim k^{- \gamma}$, $2 < \gamma < 3$,
which consists of many nodes with low degrees 
and a few hubs with high degrees.
The SF structure is quite different from the conventional simple 
regular lattices and random graphs 
(also from more complicated graphs, 
e.g. the Voronoi or Delaunay diagrams \cite{Okabe00}), 
and extremely vulnerable to 
intentional attacks on hubs \cite{Albert00}.
By removing only about a few percent of high-degree nodes, 
a set of disconnected nodes leads to 
the global malfunction of a communication network \cite{Satorras04}.
Thus, 
the design of more robust networks than the SF structure,
such as in the Internet or a Peer-to-Peer system, 
is important to reduce the threat of cyber-terrorism and of a natural
disaster to communication infrastructures.

Recently, 
it has been analytically and numerically 
shown \cite{Tanizawa06} that 
the robustness against both random and targeted removals of nodes 
is improved as the modality of degree is smaller
in the multimodal networks with a specific type of degree distribution 
$P(k_{i}) \propto a^{-(i-1)}$, $a > 1$, 
$i = 1, 2, \ldots, m$: maximum modality, 
which include the SF structure as the largest modality
with $m \rightarrow \infty$. 
Although the bimodal network with only two types of degree 
is the optimal in this class of networks under the assumption of
uncorrelated tree-like structure,
we can consider any other types of degree. 
Even with a small modality, 
the best allocation of degrees to nodes is generally unknown, 
and it may depend on the geographical positions of nodes. 
Clearly, in real communication networks, 
the position of node is not uniformly distributed 
\cite{Yook02} 
owing to 
the preference of crowded urbanism or geographical limitations on
residences.
Therefore, the spatial distribution is non-Poisson. 

On the other hand, 
point processes give useful theoretical insights for the modeling of 
spatial distribution of nodes in wireless communication networks.
Some mathematical approaches have been developed 
\cite{Blazsczyszy04}, 
while studying models other than e.g. 
the Poisson Voronoi tessellations and the Gibbs point process 
for a decomposition into some territories or 
other tessellation model for crack patterns \cite{Nagel07}
remains in an open and potential research field.
In a different point process from the above models,
we focus on the robustness of connectivity 
and on the efficiency of routing inspired by the progress 
in computer science and in 
complex network science \cite{Barabasi02,Buchanan02}.
The topological structure and the traffic dynamics based on various routing
protocols 
on complex networks have been studied actively to avoid the
congestion of packets  
(e.g. see \cite{Zhao5,Yan06,Wang06,Sreen07}).

In addition, 
we are motivated by some geometric constructions of 
spatially grown SF networks \cite{Doye05,Zhou05}, 
in which newly added nodes and links are determined by the 
positions of already existing nodes. 
Our optimal policy is different from that in the growing complex
networks by local linkings on a space \cite{Zhou07,Brunet07}
in which the emergence of various topological
structures is mainly discussed instead of the robustness of connectivity 
and the efficiency of routing.
In computer science, 
geometric network models 
(e.g. Gabriel \cite{Kranakis06}
and $\Theta$-graphs \cite{Farshi05} 
in restricted power consumption)
have also been proposed. 
Most of the research has been devoted to algorithmic 
and graph theoretical issues \cite{Li03} about efficient 
routing and economical linking on a general 
(usually, uniformly random) position of node.
However,
percolation analysis in statistical physics 
should be included in the discussion, 
because the robust infrastructure 
on a realistic space and the routing scheme
are closely related \cite{Kranakis06} in the design of network. 
The positions of nodes and the linkings strongly affect the 
distances of optimal paths and the tolerance of connectivity.
Thus, by including the issues of routing and robustness, 
we consider how to design a communication network 
in realistic positions of nodes as base stations
and linkings between them 
on a heterogeneously distributed population.

The organization of this paper is as follows. 
In Section \ref{sec2}, we introduce a new network construction 
according to a given distribution of 
a spatially heterogeneous population. 
We consider an incremental design of a communication network 
and a good property for decentralized routings. 
In Section \ref{sec3}, 
we numerically investigate 
the robustness of connectivity against both random failures and the
intentional attacks on high degree nodes in 
the geographical networks. 
In particular, we point out a more weakened tolerance by geographical
constraints on local cycles than the
random null model under the same degree distribution, 
and propose a practical strategy to improve the tolerance 
by adding a small fraction of shortcuts. 
In Section \ref{sec4}, 
we summarize the results and 
briefly discuss the future studies.

\section{Geographical network based on a population} \label{sec2}
We propose 
an evolutionary construction of geographical networks
with good properties of small modality of degree, 
short distance path, and decentralized routing. 
The robustness of connectivity will be discussed in the later sections.

\subsection{Geometric spanner} 
Let us consider the basic process of network evolution defined as follows.
It is based on a {\em point process} for load balancing of 
incremental communication requests 
by stochastic subdivisions of a triangle. 
Each node of the triangle corresponds to 
a base station for transferring messages 
and the link between them corresponds to a 
wireless or wired communication line, however
the technical details to distinguish them at the physical device level 
is beyond our current scope of network modeling.

\begin{figure}
  \begin{minipage}[htb]{.47\textwidth}
    \begin{center}
     \includegraphics[width=62mm]{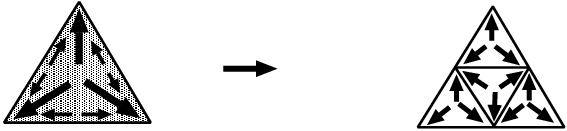}
    \end{center}
  \begin{center} (a) Re-assignment to nodes \end{center}
  \end{minipage}
  \hfill
  \begin{minipage}[htb]{.47\textwidth} \vspace{3mm}
    \begin{center}
     \includegraphics[width=70mm]{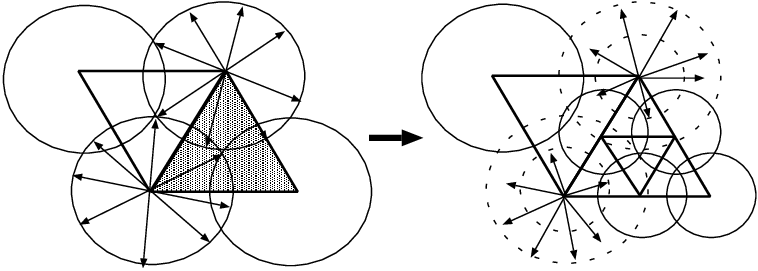}
    \end{center}
  \begin{center} (b) Regulation of communication range \end{center}
  \end{minipage}
  \caption{Basic division process. 
  (a) Re-assignment of communication requests 
  to nodes as the nearest base stations. 
  The arrows indicate the directions for
  transfered requests from users in the area. 
  (b) Power of orientative wireless beam
  regulated by the subdivision. 
  Each circle represents the range.} \label{fig_re-assign}
\end{figure}

\begin{figure}
  \begin{minipage}[htb]{.47\textwidth}
    \begin{center}
     \includegraphics[width=70mm]{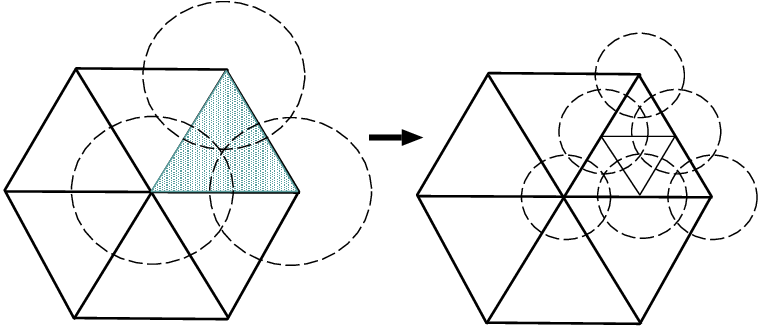}
    \end{center}
  \begin{center} (a) Subdivision \end{center}
  \end{minipage}
  \hfill
  \begin{minipage}[htb]{.47\textwidth} \vspace{3mm}
    \begin{center}
     \includegraphics[width=70mm]{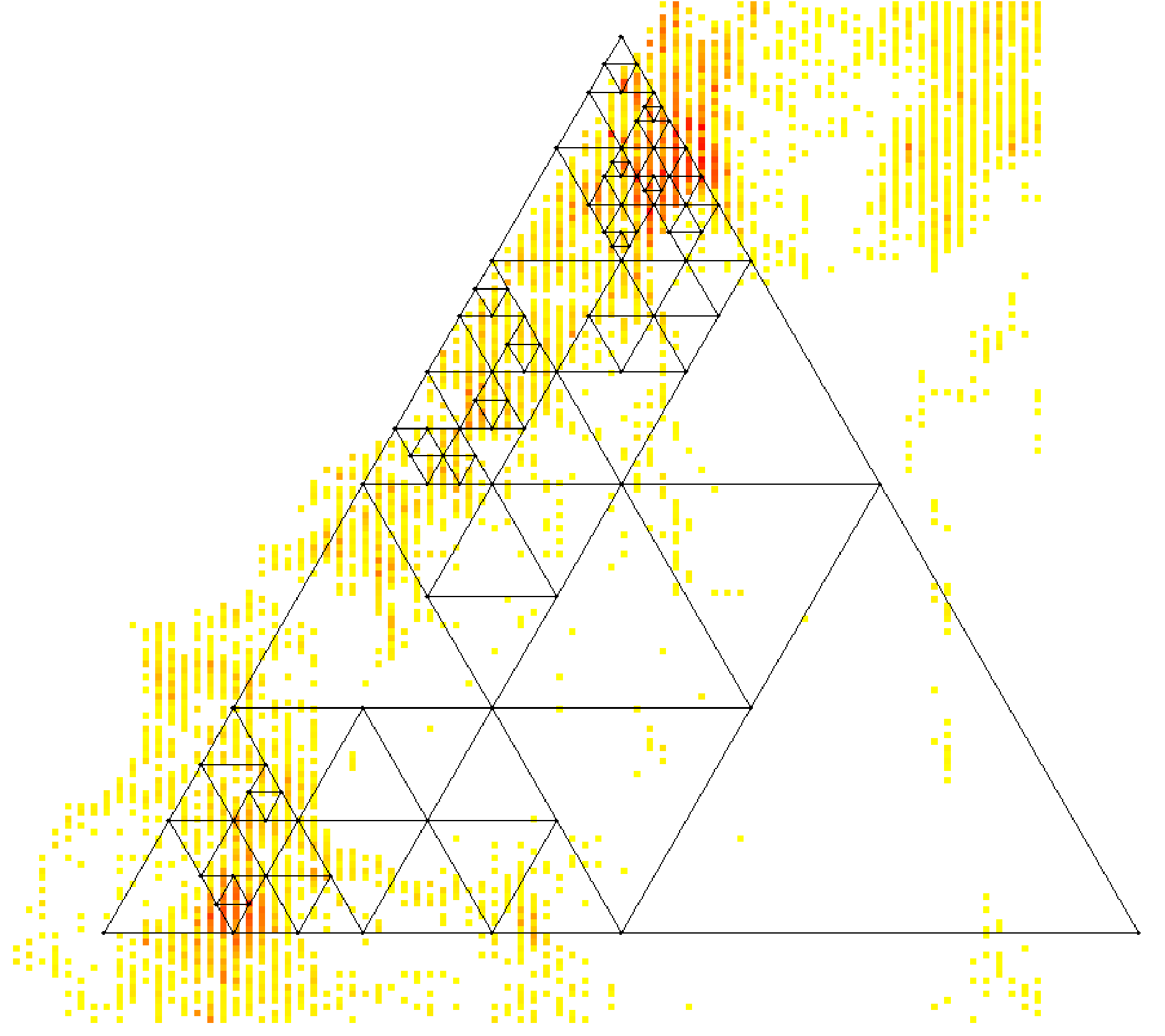}
    \end{center}
  \begin{center} (b) Example  \end{center}
  \end{minipage}
  \caption{(Color online) Heterogeneous network configuration.
  (a) Subdivision from a shaded triangle in an initial hexagon at left 
   into four smaller triangles at right. 
   The dashed circle represents the range of a
   wireless beam from each node.
  (b) Example of a network on the Fukui-Kanazawa area in Japan. 
  From light (yellow) to dark (red) color, 
  the gradation is proportionally assigned to the 
  population in each block.
  Note that the white areas at
  the upper left and the lower right are the sea of Japan 
  and the Hakusan mountain range, respectively.} \label{fig_subdivision}
\end{figure}

\begin{description}
  \item[Step0: ] Set an initial triangulation of any polygonal region
	     which consists of equilateral triangles. 
  \item[Step1: ] At each time step, a triangle is chosen with a 
	     probability proportional to 
	     the population (by summing up the number of people) 
	     in the corresponding space. 
  \item[Step2: ] Then, as shown in Figs. \ref{fig_re-assign}(a)
	     and \ref{fig_subdivision}(a), 
	     four smaller triangles are created 
	     by adding facility nodes at the intermediate points on 
	     the communication links of the chosen triangle. 
	     This procedure is for a division of the service area. 
  \item[Step3: ] Return to Step 1.
\end{description}

In the subdivision of this network, we use 
a mesh data of population statistics 
($8^{2} \times 10^{2} \times 4 = 25600$ blocks 
for $80 km^{2}$ in the Fukui-Kanazawa area, 
provided by Japan Statistical Association),
and recalculate the mapping between the blocks and each 
triangle to count the number of people 
in the triangle space.
It is natural that 
the amount of communication requests depends on 
the activities of people, 
therefore are estimated to be proportional to the 
population in the area.
We consider such a stochastic construction of geographical networks 
for the general discussion, 
however a deterministic evolutionary construction is also possible 
when the population inside the corresponding 
triangle locally accesses a certain threshold. 
We mention that, in the meaning of practical construction, 
the subdivision is generated in a distributed manner
according to the increase of communication requests in individual areas. 
The detailed procedures depend on technological issues in wide-area 
wireless communications.

According to the population, 
an example of randomly constructed network with 
a total number of nodes $N = 100$
is shown in Fig. \ref{fig_subdivision}(b). 
This configuration resembles a heterogeneous random version of 
 a Sierpinski gasket.
For any population, 
since it has a small modality with 
the trimodal degrees: $k_{1} = 2$ (or $3$ for the initial hexagon), 
$k_{2} = 4$, and $k_{3} = 6$ grown from 
the initial triangle, 
a highly tolerant connectivity may be 
expected from Ref.\cite{Tanizawa06} 
because of the small modality of degree.
Note that the smallest degree $k_{1}$ 
is fixed on the initial nodes with the existing probability 
$p_{1} = 3 / N$ (or $6 / N$ for the initial hexagon) 
in this rule of geographical network generation, 
while in the optimal bimodal networks,
$k_{1}$ with the probability 
$p_{1} = 1 - p_{2}$ is uniformly determined by 
$k_{2} = \sqrt{\langle k \rangle N}$, 
$p_{2} = \left( \frac{A^{2}}{\langle k \rangle N} \right)^{3/4}$
and 
$A =  \{ \frac{2 \langle k \rangle^{2} (\langle k \rangle - 1)^{2}
}{2 \langle k \rangle - 1} \}^{1/3}$
under the assumption of an 
uncorrelated tree-like structure of large random networks
for $N \gg 1$, then $k_{2} \gg 6$ is obtained. 
Here, $\langle k \rangle$ denotes a given average degree.
Since the differences are not only the number of modalities but also 
the size of largest degree, 
the robustness should be carefully discussed.
We will numerically  compare the robustness 
in the optimal bimodal networks 
with that in our proposed geographical networks 
under the same $\langle k \rangle$ 
at the end of next section.

On the other hand, 
as a good graph property known in computer science, 
the proposed network becomes the {\em $t$-spanner} \cite{Karavelas01} 
with a maximum stretch factor $t = 2$, 
since the equilateral property holds in spite of 
various sizes of the triangle. 
In other words, 
the network consists of the fattest triangles, 
then the length of the shortest distance path between any nodes 
$u$ and $v$ is bounded by 
$t$ times the direct Euclidean distance $d(u, v)$. 
Since 
narrow triangles as in random Apollonian networks \cite{Zhou05} 
constructed by different geometric subdivisions from ours 
give  rise to a long distance path, 
such a construction does not provide a suitable topology for 
routing paths with as short distances as possible. 
In other geometric graphs, the stretch factor becomes larger: 
$t = 2 \pi / (3 cos(\pi / 6)) \approx 2.42$ 
for Delaunay triangulations \cite{Keil92} 
and 
$t = 2 \alpha \geq 4 \sqrt{3} / 3 \approx 2.3094$ 
for two-dimensional 
triangulations with an aspect ratio of hypotenuse/height 
less than $\alpha$ \cite{Kranakis06}
whose lower bound is given for the triangulation lattice 
that consists of the fattest equilateral triangles. 
Although 
$\Theta$-graphs \cite{Farshi05} 
with $K$ non-overlapping cones have 
$t = 1 / (\cos(2 \pi / K) - \sin(2 \pi / K)) \rightarrow 1$ 
asymptotically as $K \rightarrow \infty$, 
a large amount of $O(K N)$ links is necessary, and 
the links may be crossed (as non-planar) and give rise to interferences
between wireless beams.
In general graphs, 
even the existence of a bounded stretch factor 
is uncertain.

\begin{figure}[htb]
 \begin{center}
  \includegraphics[width=137mm]{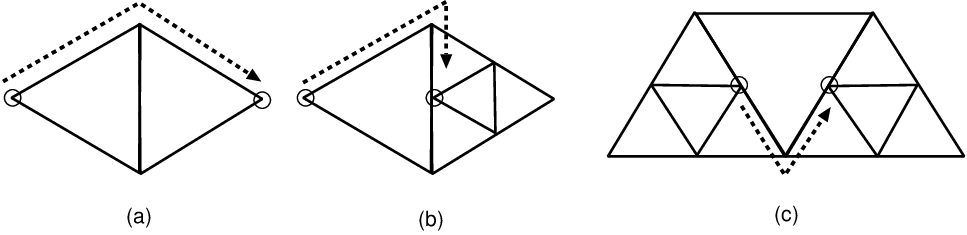}
 \end{center}
  \caption{Typical cases of the stretch factor:  
  (a) $t = 2 \sqrt{3} / 3 \approx 1.15$, 
  (b) $t = \sqrt{3} \approx 1.73$, 
  and the maximum (c) $t = 2$.
  The dashed line is the shortest distance path between 
  the source and the terminal nodes marked by circles.
  Note that all the triangles of different sizes are equilateral
  in the configuration.} 
  \label{fig_stretch}
\end{figure}

\begin{figure}
  \begin{minipage}[htb]{.47\textwidth}
    \resizebox{70mm}{!}{
    \includegraphics{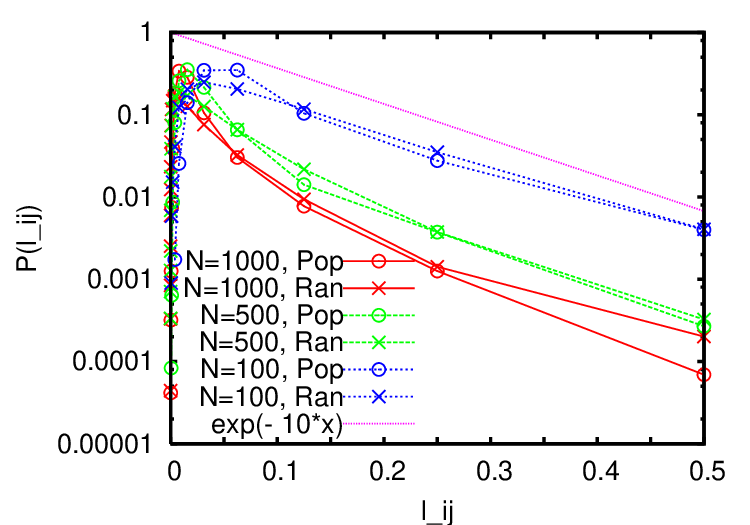}}
  \begin{center} (a) Dist. of Link Length \end{center}
  \end{minipage}
  \hfill
  \begin{minipage}[htb]{.47\textwidth}
    \resizebox{70mm}{!}{
    \includegraphics{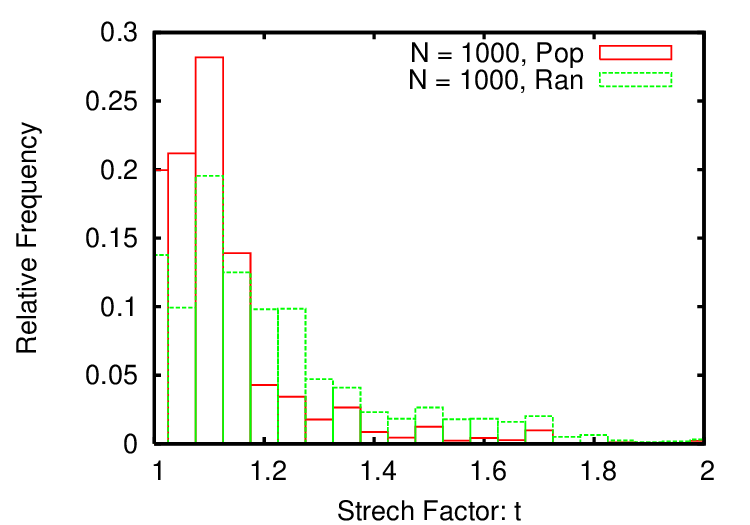}}
  \begin{center} (b) Histogram of t-value \end{center}
  \end{minipage}
  \caption{(Color online) Distance properties of the geographical 
  networks with sizes $N = 100 \sim 1000$ according to 
  the population in Fig. \ref{fig_subdivision}(b) 
  and the uniformly random choice of a triangle 
  corresponded to the case of a general population. 
  These cases are denoted as abbreviations by Pop and Ran, respectively. 
  (a) The distribution $P(l_{ij})$ of the link length $l_{ij}$ 
  decays nearly exponentially, as shown by 
  the magenta line defined by $\exp(- 10 \; l_{ij})$. 
  (b) High frequencies are observed for small stretch factors $t < 1.2$, 
  while low frequencies for large ones which are bounded by $t = 2$.}
  \label{fig_t-spanner}
\end{figure}

Figure \ref{fig_stretch} shows typical cases of the stretch factor 
on the shortest distance path in our model. 
We numerically investigate the distributions of the link length and 
of the stretch factor as shown in Fig. \ref{fig_t-spanner}. 
Many paths connected by the majority of short links 
have small stretch factors. 
The average link lengths $\bar{l}_{ij}$ are less than half of 
the initial link length normalized with refer to 
the biggest triangle, 
as $0.057$  for $N = 100$ and $0.014$ for $N = 1000$ in the Pop 
($\bar{l}_{ij} = 0.05$ for $N = 100$ 
and $0.01$ for $N = 1000$ in the Ran), 
the average factors $\bar{t}$ are small as 
$1.1512$ for $N = 100$ and $1.1395$ for $N = 1000$ 
 ($\bar{t} = 1.2008$ for $N = 100$ 
and $1.2297$ for $N = 1000$ in the Ran). 
Thus, our proposed networks have good properties regarding path 
distance.

\newpage
\subsection{Decentralized face routing}
For ad-hoc networks, a routing protocol should be simple to reduce
energy consumption in keeping the reachability. 
However some of the routing schemes in early work lead to the failure of 
guaranteed delivery \cite{Urrutia02}, 
e.g., in the flooding algorithm multiple redundant
copies of a message are sent and cause network congestion, while 
greedy and compass routings may occasionally fall into infinite loops. 
On the other hand, except for some class of graphs, 
it is costly to seek the shortest path for a map in which 
connections are unknown in advance.

\begin{figure}[htb]
 \begin{center}
  \includegraphics[width=80mm]{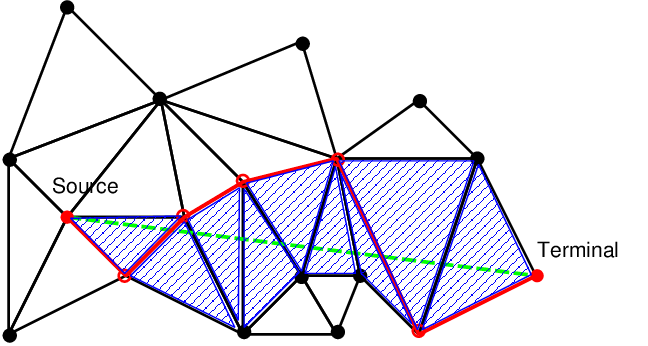}
 \end{center}
  \caption{(Color online) Efficient face routing on a planar network. 
  The (red) reachable shortest path 
  through the nodes marked by open circles 
  between the source and the terminal nodes
  can be found from the edges of the (blue) shaded faces that intersect 
  the (green) dashed straight line.
  The two paths on the edges above and below the line 
  are the upper and the lower chains, respectively.} 
  \label{fig_face_routing}
\end{figure}

Fortunately, 
since the proposed geographical networks belong to a special class of graph 
which is planar and consists of convex (equilaterally triangular) faces, 
we can apply the 
{\em efficient decentralized algorithm} \cite{Bose04} 
that guarantees the delivery of a message 
using only local information based on the positions of 
the source, the terminal nodes, 
and the adjacencies of the current node on a path.
In recent technologies, 
the positions are measured 
by means of a GPS or other methods, 
then the reachable path to a terminal node $Te$
can be found in a proper mixing of upper and
lower chains extracted from the edges of the faces 
as shown in Fig. \ref{fig_face_routing}. 
Without multiple copies of the message, 
the next forwarding node from each 
node on the upper and lower chains is determined only by the positions of 
adjacent nodes and the distance from a source node $So$.
In this class, 
with the property of routing called competitive, 
the length of the routing path is bounded in a constant factor to 
that of the shortest path and also to the distance $d(So, Te)$ 
on the straight line because of the $t$-spanner.

This competitive algorithm \cite{Bose04} or a combination 
by greedy and face routings in asymptotically 
(at the lower bound) worst-case optimal traveling 
on a general planar network
\cite{Kuhn03} acts in a decentralized manner,
therefore a global information such as the static routing
table in the Internet's TCP/IP protocol is not necessary. 
Moreover, these algorithms can be extended 
(see the Appendix of \cite{Hayashi07} 
and a related idea \cite{Kuhn02}) to add a small fraction 
of shortcuts discussed in the next section.
We emphasize that 
such geographical protocols are not only 
successfully used for message delivery in social friendships 
\cite{Nowell05} 
but are also very promising for 
constructing wireless networks 
with a higher efficiency and scalability in a dynamic environment  
\cite{Seada06}.

\section{Numerical robustness analysis} \label{sec3}
We aim to maximize the sum of critical fractions 
$f_{T} = f_{r} + f_{t}$ to be optimal 
network against both random failures and targeted attacks on nodes with
large degrees as robust as possible on a spatially 
heterogeneous distribution of population.
Here, $f_{r}$ and $f_{t}$ denote the critical fractions of these damages
at the breaking of the giant component (GC). 

For a preliminary, we numerically estimate 
the total numbers of nodes $N(\tau)$ and links $M(\tau)$ 
at time step $\tau$ over 100 realizations of 
the geographical network. 
They grow as 
$N(\tau) = N(0) + 2.35 \tau$ and $M(\tau) = M(0) + 5.34 \tau$, 
where 
$N(0) = 6$ and $M(0) = 9$ for the initial triangle
(or $N(0) = 7$ and $M(0) = 12$ for the initial hexagon). 
Thus, in the subdivision, 
two or three nodes per step are added with a high frequency, 
and the average degree is 
$\langle k \rangle = 2 M(\tau) / N(\tau) \approx 4.54$. 
Since the total link $\langle k \rangle N / 2$ is more than 
twice that of $N-1$ links on a tree of $N$ nodes,
there possibly exist many cycles.

\begin{figure}
  \begin{minipage}[htb]{.47\textwidth}
    \resizebox{70mm}{!}{
    \includegraphics{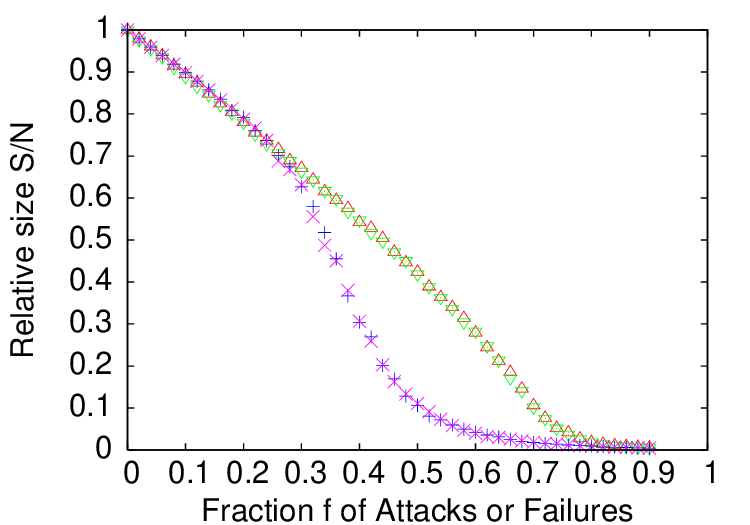}}
  \begin{center} (a) Giant Component \end{center}
  \end{minipage}
  \hfill
  \begin{minipage}[htb]{.47\textwidth}
    \resizebox{70mm}{!}{
    \includegraphics{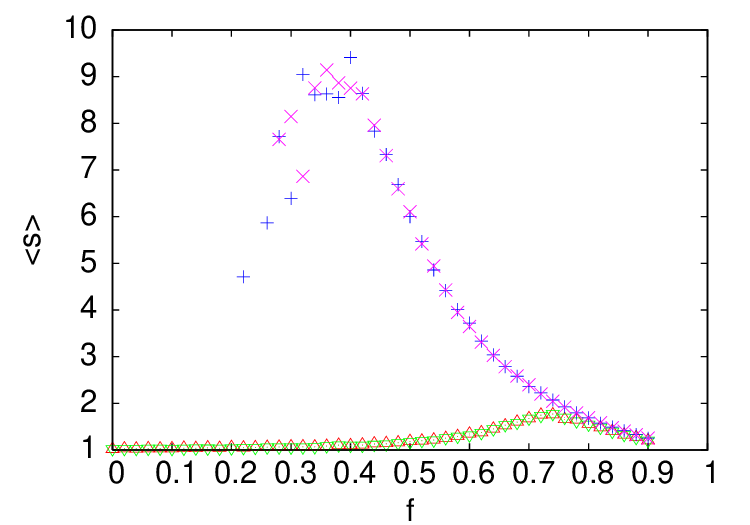}}
  \begin{center} (b) Isolated Clusters \end{center}
  \end{minipage}
  \caption{(Color online) Comparison of the geographical and the
  non-geographical rewired networks at $N = 1000$.
  (a) Relative size $S/N$ of the GC after the removal of nodes. 
  The plus(blue) and cross(magenta) marks indicate the corresponding
  results 
  for the geographical networks against random failures and intentional 
  attacks. 
  The upper triangle(red) and down triangle(green) marks indicate the
  corresponding results
  for the non-geographical rewired versions against them, respectively.
  (b) Average size $\langle s \rangle$
  of isolated clusters except the GC. 
  The peak indicates the critical point 
  at which the whole connectivity breaks. 
  Each point is obtained from the average 
  over 100 realizations of the geographical network 
  constructed from the initial triangle 
  ($\times$ 100 samples of random rewirings).} 
  \label{fig_geo_rew}
\end{figure}

As the simulation result for the removals, 
Fig. \ref{fig_geo_rew}  indicated with blue $+$ and magenta 
$\times$ marks show that 
the size $S$ of the GC 
rapidly decreases at the fraction $f \approx 0.4$.
A similar 
vulnerability caused by geographical constraints on local cycles 
has been found 
in a family of SF networks embedded in a planar space \cite{Hayashi06} 
and SF networks on a lattice \cite{Huang05}. 
To clearly see the effect of constraints,  
we investigate the non-geographical rewired versions 
under the same degree distribution 
(of course with the same average degree $\langle k \rangle$). 
In the rewiring \cite{Maslov04}, 
two pairs of nodes at the ends 
of randomly chosen links are exchanged 
in holding the degree of each node.
Therefore, the geographical constraints are entirely reduced. 
In general, 
the rewired version of a network is the null model 
that depend only on the degree distribution ignoring the other 
topological structures: cycles, degree-degree correlation, 
fractal or hierarchical substructure, 
diameter of graph, etc. 
The red $\bigtriangleup$ and green $\bigtriangledown$ marks shown in 
Fig. \ref{fig_geo_rew} indicate
that the critical fractions at the peak of $\langle s \rangle$
increases to $f_{r} \approx 0.7$ and $f_{t} \approx 0.6$. 
Such improvement of the robustness 
is consistent with the result obtained for the geographical SF networks 
\cite{Hayashi06}.

\begin{figure}
  \begin{minipage}[htb]{.47\textwidth}
    \resizebox{70mm}{!}{
    \includegraphics{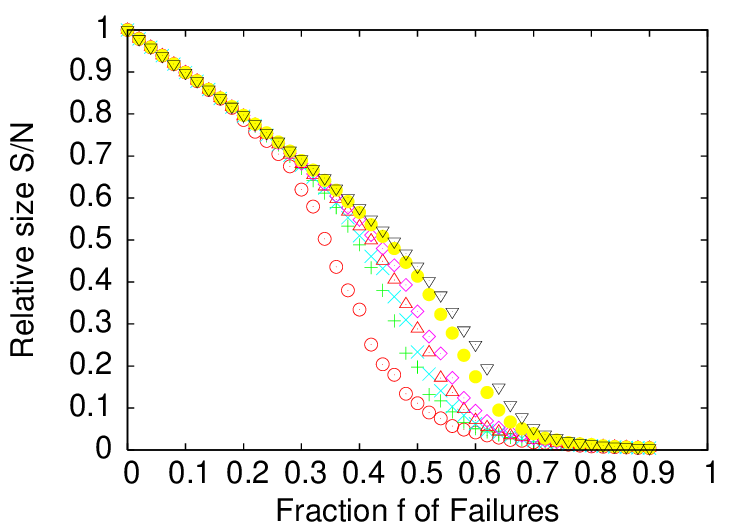}}
  \begin{center} (a) Giant Component \end{center}
  \end{minipage}
  \hfill
  \begin{minipage}[htb]{.47\textwidth}
    \resizebox{70mm}{!}{
    \includegraphics{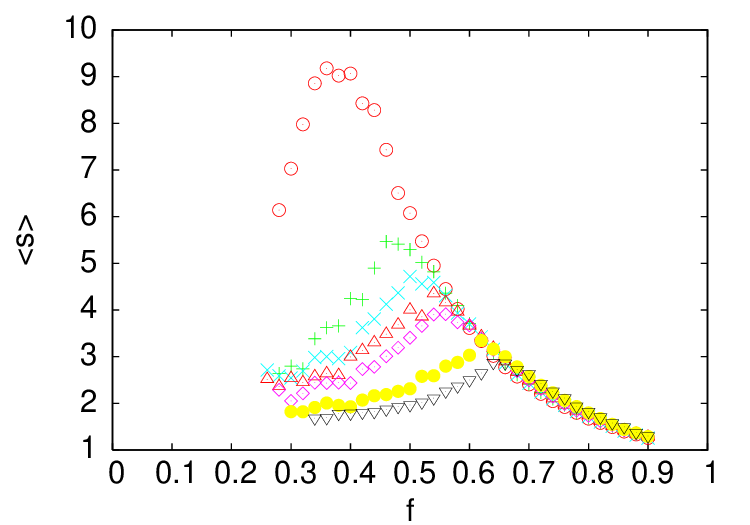}}
  \begin{center} (b) Isolated Clusters \end{center}
  \end{minipage}
  \caption{(Color online) Damages by random failures in the geographical
  networks and the shortcut versions at $N = 1000$. 
  (a) Relative size $S/N$ of the GC.
  The open circle(red), plus(green), cross(cyan), 
  upper triangle(orange), diamond(magenta), closed circle(yellow), and 
  down triangle(black) marks indicate the corresponding results for the
  shortcut rate of 0, 3, 5, 7, 10, 20, and 30 \%, respectively. 
  (b) Average size $\langle s \rangle$
  of isolated clusters except the GC. 
  Each point is obtained from the average 
  over 100 realizations of the geographical network constructed 
  from the initial triangle $\times$ 100 samples of random shortcuts.}
  \label{fig_shortcuts_rand}
\end{figure}

\begin{figure}
  \begin{minipage}[htb]{.47\textwidth}
    \resizebox{70mm}{!}{
    \includegraphics{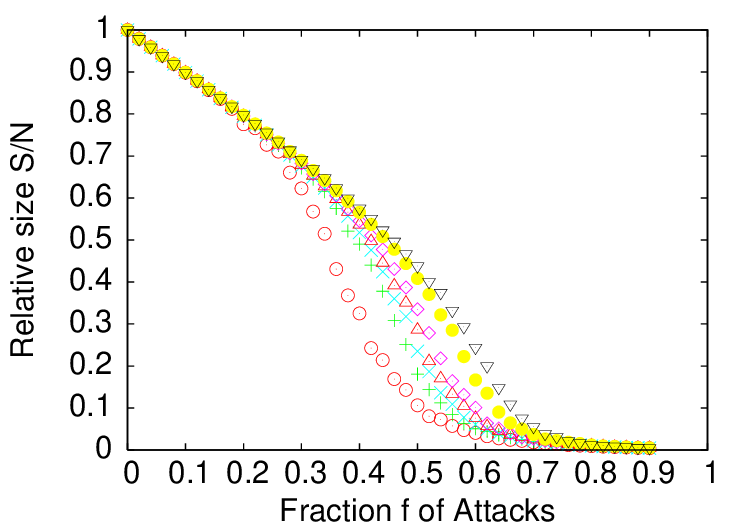}}
  \begin{center} (a) Giant Component \end{center}
  \end{minipage}
  \hfill
  \begin{minipage}[htb]{.47\textwidth}
    \resizebox{70mm}{!}{
    \includegraphics{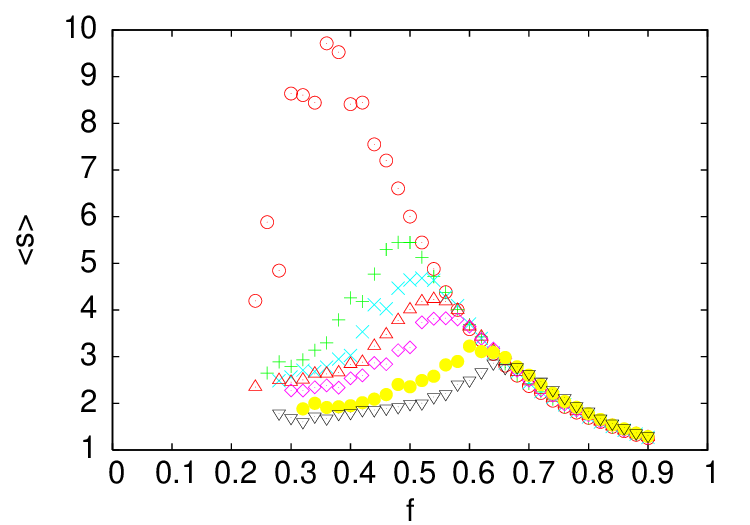}}
  \begin{center} (b) Isolated Clusters \end{center}
  \end{minipage}
  \caption{(Color online) Damages by intentional attacks in the
  geographical networks and the shortcut versions at $N = 1000$. 
  (a) Relative size $S/N$ of the GC.
  (b) Average size $\langle s \rangle$
  of isolated clusters except the GC. 
  The marks are the same as in Fig. \ref{fig_shortcuts_rand}.
  Each point is obtained from the average 
  over 100 realizations of the geographical network constructed 
  from the initial triangle $\times$ 100 samples of random shortcuts.}
  \label{fig_shortcuts_attack}
\end{figure}

Although the full rewiring is better in terms of the robustness, 
it completely ignores 
the positions of nodes and the distances of links. 
As another practical strategy, it is expected that 
adding a small fraction of shortcuts between randomly chosen nodes 
has a similar effect to the rewiring \cite{Hayashi07}. 
Indeed, 
Figs. \ref{fig_shortcuts_rand} and \ref{fig_shortcuts_attack}
show the improvement of robustness against
both random failures and intentional attacks. 
As the shortcut rate marked 
from red $\bigcirc$ to black $\bigtriangledown$ increases, 
a larger GC remains, 
and the breaking point at the peak of $\langle s \rangle$ 
shifts to a larger fraction $f$ of the removal. 
Only about 10 \% of the adding reaches 
a similar level of 
$f_{r} \approx 0.7$ or $f_{t} \approx 0.6$ in 
the non-geographical rewired version
(compare with Fig. \ref{fig_geo_rew} indicated with 
$\bigtriangleup$ and $\bigtriangledown$ marks). 
In addition, the finite effect on the critical fractions 
is very small as shown in Table \ref{table_no_finite},
however it has occurred that 
the positioning of the node is inaccurate written a round-off error in
the very dense case of $N = 10^{5}$, especially for $0 \%$ shortcuts.

\begin{figure}
  \begin{minipage}[htb]{.47\textwidth}
    \resizebox{73mm}{!}{
    \includegraphics{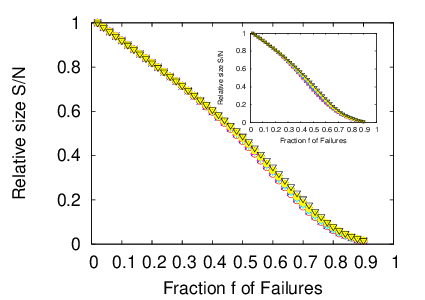}}
  \begin{center} (a) Giant Component \end{center}
  \end{minipage}
  \hfill
  \begin{minipage}[htb]{.47\textwidth} \hspace{-5mm}
    \resizebox{73mm}{!}{
    \includegraphics{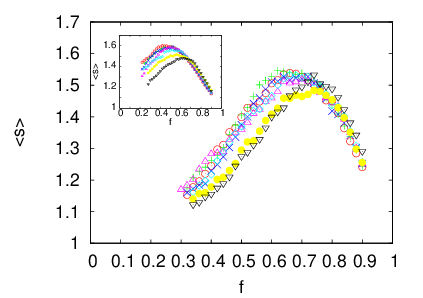}}
  \begin{center} (b) Isolated Clusters \end{center}
  \end{minipage}
  \caption{(Color online) Damages by random failures in the optimal
  bimodal networks and the shortcut versions at $N = 1000$ with 
  the over-estimated $\langle k \rangle = 4.882 > 4.54$,
  $k_{1} = 4$, and $k_{2} = 67$. 
  (a) Relative size $S/N$ of the GC. 
  (b) Average size $\langle s \rangle$
  of isolated clusters except the GC. 
  Insets show the results with the under-estimated 
  $\langle k \rangle = 3.975 < 4.54$,
  $k_{1} = 3$, and $k_{2} = 68$.
  Each point is obtained from the average 
  over 100 realizations $\times$ 100 samples of random shortcuts.
  The marks are the same as in Fig. \ref{fig_shortcuts_rand}.}
  \label{fig_bimodal_rand}
\end{figure}

\begin{figure}
  \begin{minipage}[htb]{.47\textwidth}
    \resizebox{73mm}{!}{
    \includegraphics{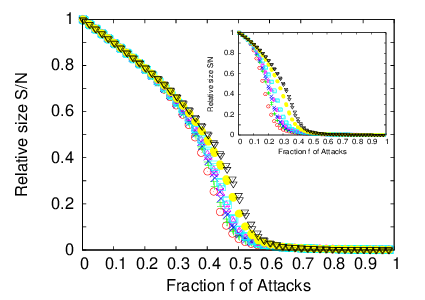}}
  \begin{center} (a) Giant Component \end{center}
  \end{minipage}
  \hfill
  \begin{minipage}[htb]{.47\textwidth} \hspace{-5mm}
    \resizebox{73mm}{!}{
    \includegraphics{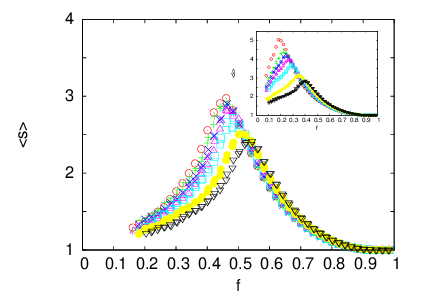}}
  \begin{center} (b) Isolated Clusters \end{center}
  \end{minipage}
  \caption{(Color online) Damages by intentional attacks in the optimal 
  bimodal networks and the shortcut versions at $N = 1000$
  with the over-estimated $\langle k \rangle = 4.882 > 4.54$, 
  $k_{1} = 4$, and $k_{2} = 67$. 
  (a) Relative size $S/N$ of the GC. 
  (b) Average size $\langle s \rangle$
  of isolated clusters except the GC. 
  Insets show the results with the under-estimated 
  $\langle k \rangle = 3.975 < 4.54$,
  $k_{1} = 3$, and $k_{2} = 68$.
  Each point is obtained from the average 
  over 100 realizations $\times$ 100 samples of random shortcuts.
  The marks are the same as in Fig. \ref{fig_shortcuts_rand}.}
  \label{fig_bimodal_attack}
\end{figure}

We compare the above results shown in Figs. \ref{fig_shortcuts_rand}
and \ref{fig_shortcuts_attack} with the robustness of 
the optimal bimodal network \cite{Tanizawa06} defined by 
$k_{2} = \sqrt{\langle k \rangle N} = 67.38$ and 
$p_{2} = \left( \frac{A^{2}}{\langle k \rangle N} \right)^{3/4} =
0.01445$
for $N = 1000$ and $\langle k \rangle = 4.54$.
Since the degree and the number of nodes for each type of degree 
must be an integer, 
the appropriate combination is $k_{2} = 67$ or $68$, 
$p_{2} = 0.014$ or $0.015$, 
$k_{1} = (\langle k \rangle - k_{2} p_{2}) / p_{1} = 4$ or $3$, 
and 
$p_{1} = 1 - p_{2} = 0.986$ or $0.985$; 
then, $\langle k \rangle = 3.975$ or $4.882$ 
is the closest to the value $4.54$
in our geographical networks. 
On the two pairs of degree distributions, 
the bimodal networks are randomly constructed by full rewirings from an 
initial configuration.
As shown in Figs. \ref{fig_bimodal_rand} and \ref{fig_bimodal_attack}, 
the critical fractions $f_{r}$ and $f_{t}$ are slightly larger 
around $0.4 \sim 0.7$, 
however the effect of shortcuts barely works 
and is weaker than that in our geographical networks.

\begin{table}[htb]
\begin{center}
\begin{small}
\begin{tabular}{c|cccc|cccc} \hline
Shortcut & & $f_{r}$ & & & & $f_{t}$ & & \\
Rate \%  & $N=10^{2}$ & $10^{3}$ & $10^{4}$ & $10^{5}$ 
         & $N=10^{2}$ & $10^{3}$ & $10^{4}$ & $10^{5}$\\ \hline
       0 & 0.352 & 0.348 & 0.386 & 0.058 & 0.351 & 0.335 & 0.384 & 0.100
 \\
       3 & 0.474 & 0.443 & 0.499 & 0.469 & 0.469 & 0.481 & 0.425 & 0.468
 \\
       5 & 0.529 & 0.501 & 0.542 & 0.502 & 0.509 & 0.499 & 0.540 & 0.503
 \\
       7 & 0.525 & 0.537 & 0.554 & 0.538 & 0.520 & 0.525 & 0.503 & 0.535
 \\
      10 & 0.545 & 0.542 & 0.566 & 0.568 & 0.568 & 0.552 & 0.571 & 0.565
 \\
      20 & 0.600 & 0.605 & 0.627 & 0.639 & 0.619 & 0.596 & 0.635 & 0.638
 \\
      30 & 0.646 & 0.641 & 0.671 & 0.682 & 0.651 & 0.641 & 0.663 & 0.680
 \\ \hline
Rewired  & 0.726 & 0.722 & 0.703 & 0.687 & 0.718 & 0.736 & 0.699 & 0.690 \\ \hline
\end{tabular}
\end{small}
\end{center}

\vspace{2mm}
\caption{Critical fractions $f_{r}$ and $f_{t}$ at the peak of 
$\langle s \rangle$ are consistent for different sizes 
$N$ of the geographical networks with shortcuts and the rewired
versions. Each value is obtained from the average 
over 100 realizations of the geographical network constructed 
from the initial triangle $\times$ 100 samples of 
random shortcuts and rewirings.}
\label{table_no_finite}
\end{table}

\section{Conclusion} \label{sec4}
According to spatially distributed communication requests based on a
given population density, 
we have proposed an evolutionarily constructed geographical network
by the iterative division of equilateral triangles. 
Through the numerical simulation for investigating the robustness, 
the obtained results are summarized as follows. 
We note that they are consistent with other results on 
population densities 
besides the example of Fig. \ref{fig_subdivision}(b), 
e.g. for the the geographical networks generated by the 
iterative choice
of a triangle with a uniformly random probability,
the robustness is similar to that 
in Figs. \ref{fig_geo_rew}-\ref{fig_shortcuts_attack}.

\begin{itemize}
  \item The proposed networks have suitable
	properties of short paths as the $t$-spanner \cite{Karavelas01}
	and efficient decentralized routing \cite{Bose04} 
	for wireless communications. Moreover, the incremental 
	construction can be implemented to accommodate a 
	growing activity or population.
  \item To improve the vulnerability 
	caused by the geographical 
	constraints \cite{Hayashi06,Huang05}, 
	we have considered a practical strategy 
	by adding a small fraction of shortcuts 
	between randomly chosen nodes \cite{Hayashi07}, 
	and numerically confirmed the effect. 
  \item The degree distribution becomes trimodal at most
	without hub nodes, 
	the robustness of the connectivity is slightly weak 
	but is maintained as 
	similar level as the optimal bimodal networks 
	with a larger maximum degree \cite{Tanizawa06}.
\end{itemize}
These results are useful for the design of ad-hoc networks with 
efficiency, scalability, and tolerance of connectivity
in wide-area communications.

Since our trimodal model has 
a maximum degree that is not extremely large compared to other degrees, 
instead of hub attacks, 
a spatial cutting into several dense parts 
by removing a small number of nodes on lines of large triangles 
may be considerable. 
Even in such cases, we expect that 
shortcuts effectively work to bridge isolated clusters by the removals.
More detailed analysis is the subject of future study that includes 
how to find the structural vulnerable points.
Other important issues are the 
development of routing schemes taking into account the structure
layered by the size of the triangle, 
the analysis of traffic dynamics with the phase
transition between free flow and congestion according to the forwarding
capacity at a node and queue discipline (FIFO etc.), 
and considering various optimal policies in matching 
application requirements.

\section*{Acknowledgments}
The author would like to thank anonymous reviewers for valuable comments 
to improve the manuscript, and also Yasumasa Ono and Hironori Okumura 
in my laboratory for helping with the simulation.
This research is supported in part by
Grant-in-Aid for Scientific Research in Japan No.18500049.


\begin{thebibliography}{50}
\bibitem{Yook02} S.-H. Yook,
H. Jeong, and A.-L. Barab\'{a}si,
{\em PNAS}, {\bf 99(21)}, 13382, (2002).

\bibitem{Gastner06} M.T. Gastner, and M.E.J. Newman,
{\em Eur. Phys. J. B}, {\bf 49(2)}, 247, (2006).


\bibitem{Okabe00} A. Okabe, B. Boots, K. Sugihara, and S.N. Chiu,
Spatial Tessellations, 2nd ed., 
John Wiley, 2000.

\bibitem{Albert00}  R. Albert, and A.-L. Barab\'{a}si,
{\em Nature}, {\bf 406}, 378, (2000).

\bibitem{Satorras04} R. Pastor-Satorras, and A. Vespignani, 
Evolution and Structure of the Internet, 
Cambridge University Press, 2004.

\bibitem{Tanizawa06}
T. Tanizawa, G. Paul, S. Havlin, and H.E. Stanley,
{\em Phys. Rev. E}, {\bf 74}, 016125, (2006).

\bibitem{Blazsczyszy04} B. Blazsczyszyn, and R. Schott, 
{\em Japan J. Ind. Appl. Math.}, {\bf 22(2)}, 179, (2005).

\bibitem{Nagel07} W. Nagel, J. Mecke, J. Ohser, and V. Weiss,
The 12th Int. Congress for Stereogy, (2007).
http://icsxii.univ-st-etiene.fr/Pdfs/f14.pdf

\bibitem{Barabasi02} A.-L. Barab\'{a}si,
LINKED: The New Science of Networks, 
Perseus, Cambridge, MA, 2002.

\bibitem{Buchanan02} M. Buchanan,
NEXUS: Small Worlds and the Groundbreaking Science of Networks,
W.W.Norton, New York, 2002.


\bibitem{Zhao5}
L. Zhao, Y.-C. Lai, K. Park, and N. Ye,
{\em Phys. Rev. E}, {\bf 71}, 026125, (2005).

\bibitem{Yan06}
G. Yan, T. Zhou, B. Hu, Z.-Q. Fu, and B.-H. Wang, 
{\em Phys. Rev. E}, {\bf 73}, 046108, (2006).

\bibitem{Wang06}
W.-X. Wang, B.-H. Wang, C.-Y. Yin, Y.-B. Xie, and T. Zhou, 
{\em Phys. Rev. E}, {\bf 73}, 026111, (2006).

\bibitem{Sreen07}
S. Sreenivasan, R. Cohen, E. L\'{o}rez, Z. Toroczkai, 
 and H.E. Stanley, 
{\em Phys. Rev. E}, {\bf 75}, 036105, (2007).



\bibitem{Doye05}
J.P.K. Doye, and C.P. Massen, 
{\em Phys. Rev. E}, {\bf 71}, 016128, (2005).

\bibitem{Zhou05}
T. Zhou, G. Yan, and B.-H. Wang,
{\em Phys. Rev. E}, {\bf 71}, 046141, (2005).

\bibitem{Zhou07}
Y.-B. Xie, T. Zhou, W.-J. Bai, G. Chen, W.-K. Xiao, and B.-H. Wang, 
{\em Phys. Rev. E}, {\bf 75}, 036106, (2007).

\bibitem{Brunet07}
R. Xulxi.-Brunet, and I.M. Sokolov, 
{\em Phys. Rev. E}, {\bf 75}, 046117, (2007).

\bibitem{Kranakis06}
E. Kranakis, and L. Stacho,
In {\em Handbook of Algorithms for Wireless Networking and Mobile
	Computing},
edited by A. Boukerche,
(Chapman \& Hall/CRC, 2006), Chap. 8.

\bibitem{Farshi05}
M. Farshi, and J. Gudmundsson, 
{\em Proc. of the 13th European Symposium on Algorithms}, 
edited by G.S. Brodal and S. Leonardi, ESA 2005, 
LNCS, {\bf 3669}, 556, (2005).


\bibitem{Li03}
X.-Y. Li,
{\em Wireless Computing and Mobile Computing}, {\bf 3}, 119, (2003).

\bibitem{Karavelas01}
M.I. Karavelas, and L.J. Guibas, 
{\em Proc. of the 12th ACM-SIAM Symposium 
on Discrete Algorithms}, (2001).

\bibitem{Keil92}
J.M. Keil, and C.A. Gutwin, 
{\em Discrete and Computational Geometry}, {\bf 7}, 13, (1992).

\bibitem{Urrutia02}
J. Urrutia,
In {\em Handbook of Wireless Networks and Mobile Computing},
edited by I. Stojmenovi\'{c},
(John Wiley \& Sons, 2002), Chap. 18.

\bibitem{Bose04}
P. Bose, and P. Morin,
{\em Theoretical Computer Science}, {\bf 324(2-3)}, 273, (2004).


\bibitem{Kuhn03}
F. Kuhn, R. Wattenhofer, and A. Zollinger,
{\em Proc. of the 4th ACM International Workshop on Mobile ad hoc
	Networking and Computing}, (MObiHOc'03), 
pp.267-278, (2003), 

\bibitem{Kuhn02}
F. Kuhn, R. Wattenhofer, and A. Zollinger,
{\em Proc. of the 6th International Workshop on Discrete Algorithms and
	Methods for Mobile Computing and Communications}, 
(DialM'02), pp.24-33, (2002).

\bibitem{Hayashi07}
Y. Hayashi, and J. Matsukubo, 
{\em Physica A}, {\bf 380}, 552, (2007).

\bibitem{Nowell05} D. Liben-Nowell, J. Novak, R. Kumar, 
P. Raghaven, and A. Tomkins,
{\em PNAS}, {\bf 102}, 11623, (2005).


\bibitem{Seada06}
K. Seada, and A. Helmy, 
In {\em Handbook of Algorithms for Wireless Networking and Mobile
	Computing},
edited by A. Boukerche,
(Chapman \& Hall/CRC, 2006), Chap. 15.

\bibitem{Hayashi06}
Y. Hayashi, and J. Matsukubo, 
{\em Phys. Rev. E}, {\bf 73}, 066113, (2006).

\bibitem{Huang05}
L. Huang, L. Yang, and K. Yang, 
{\em Europhys. Lett.}, {\bf 72(1)}, 144, (2005).


\bibitem{Maslov04} S. Maslov, K. Sneppen, and A. Zaliznyak,
{\em Physica A} {\bf 333}, 529, (2004).

\end{thebibliography}
\end{document}